\documentclass{conm-p-l}

\copyrightinfo{2015}{}

\usepackage{graphicx}
\usepackage{subfigure}

\usepackage{amssymb,amsmath,amsthm,amscd}

\theoremstyle{definition}

\theoremstyle{remark}

\numberwithin{equation}{section}

\newcommand{\ra}{\rightarrow}

\newcommand{\sg}{\sigma}
\newcommand{\Sg}{\Sigma}

\newcommand{\La}{\Lambda}

\renewcommand{\O}{{\mathcal O}}

\def\maprightu#1{\smash{
    \mathop{\longrightarrow}\limits^{#1}}}
\def\maprightd#1{\smash{
    \mathop{\longrightarrow}\limits_{#1}}}
\def\mapdownl#1{
    \llap{$\vcenter{\hbox{$\scriptstyle#1$}}$}\Big\downarrow}
\def\mapdownr#1{\Big\downarrow
    \rlap{$\vcenter{\hbox{$\scriptstyle#1$}}$}}

\begin{document}

\title[Symbolic dynamics and chaos in plane Couette flow]{Symbolic dynamics and chaos 
in plane Couette flow}

\author{Y. Charles Li}
\address{Department of Mathematics, University of Missouri, 
Columbia, MO 65211, USA}
\email{liyan@missouri.edu}
\urladdr{http://www.math.missouri.edu/~cli}

\curraddr{}
\thanks{}

\date{}

\dedicatory{}


\begin{abstract}
According to a recent theory \cite{Li14}, when the Reynolds number is large, fully developed 
turbulence is caused by short term unpredictability (rough dependence upon initial data); when 
the Reynolds number is moderate, often transient turbulence is caused by chaos (long term 
unpredictability). This article aims at studying chaos in plane Couette flow at moderate Reynolds 
number. Based upon the work of L. van Veen and G. Kawahara \cite{VK11} on a transversal homoclinic orbit
asymptotic to a limit cycle in plane Couette flow, we explore symbolic dynamics and chaos near the 
homoclinic orbit. Mathematical analysis shows that there is a collection of orbits in the neighborhood 
of the homoclinic orbit, which is in one-to-one correspondence with the collection 
of binary sequences. The Bernoulli shift on the binary sequences corresponds to a chaotic dynamics 
of a properly defined return map. 
\end{abstract}

\maketitle

\section{Introduction}

From our analytical result \cite{Li14} (supported by our numerical experiments \cite{LL15}), perturbations in turbulence amplify in time according to $e^{\sg \sqrt{Re} \sqrt{t} + \sg_1 t}$ where $\sg_1 = \frac{\sqrt{2e}}{2} \sg$ and $\sg$ depends only on the base solutions on which the perturbations are introduced. When the time is small, the first term in the exponent dominates, and this term can cause the amplification to be superfast when the Reynolds number is large. By the time $t \sim Re$, the two terms in the exponent are about equal. After the time $t \sim Re$, the second term dominates, and this term is the classical Liapunov exponent that causes chaos (long term unpredictability). Thus the time $t \sim Re$ is the temporal separation point between short term unpredictability and long term unpredictability. When the Reynolds number is large, long before the separation point $t \sim Re$, the first term in the exponent already amplifies the perturbation to substantial size, and the second term does not get a chance to act. Thus fully developed turbulence is dominated by such short term unpredictability.  When the Reynolds number is moderate, both terms in the exponent have a chance to dominate, and the corresponding (often) transient turbulence is dominated by chaos in long term. Since random perturbations can happen at any time, this leads to the fact that the exact mathematical solution of the Navier-Stokes equations does not describe any turbulent flow even approximately in short time, when the Reynolds number is large. This is in sharp contrast to chaos for which the exact mathematical solution at least approximates a chaotic flow in finite time. We study the large Reynolds number Navier-Stokes equations somewhere else \cite{Li15}. Here we study moderate Reynolds number Navier-Stokes equations and focus upon symbolic dynamics and chaos near the homoclinic orbit discovered by L. van Veen and G. Kawahara \cite{VK11}. 
There have been a lot of dynamical system studies on moderate Reynolds number Navier-Stokes equations, and various dynamical system objects have been discovered \cite{KUV12}, e.g. fixed points, periodic orbits, periodic doubling bifurcation \cite{KE12} \cite{ZE15}, heteroclinic tangle \cite{ME12}. These studies were conducted on specific flows like plane Couette flow, plane Poiseuille flow, and pipe Poiseuille flow. 

\section{Bernoulli shift on binary sequences}

The Bernoulli shift is an automorphism that has chaotic dynamics. In fact, it is a canonical chaos map to which other chaos maps can be shown to be topologically conjugate to. Let $\Sg$ be the set of all doubly infinite binary sequences:
\[
a = ( \cdots a_{-2} a_{-1} a_0,a_1a_2 \cdots ),
\]
where $a_n = 0$ or $1$ for any integer $n$. For any point in $\Sg$,
\[
a^* = ( \cdots a^*_{-2} a^*_{-1} a^*_0,a^*_1a^*_2 \cdots ),
\]
one can define the neighborhood basis
\[
U_j(a^*) = \left \{  a \in \Sg \ | \ a_n = a^*_n , \ |n| < j \right \}
\]
where $j = 1, 2, \cdots $.  These $U_j(a^*)$'s generate a topology for $\Sg$. Endowed with this topology, $\Sg$ becomes a topological space. The larger the $j$ is, the smaller the neighborhood 
$U_j(a^*)$ is. The Bernoulli shift $\chi$ is a map defined in $\Sg$. For any point $a$ in $\Sg$, let $b = \chi (a)$, then $b_n = a_{n+1}$ for all integer $n$. That is, if $a$ is given by
\[
a = ( \cdots a_{-2} a_{-1} a_0,a_1a_2 \cdots ),
\]
then $\chi (a)$ is given by
\[
\chi (a) = ( \cdots a_{-2} a_{-1} a_0a_1,a_2 \cdots ),
\]
in other words, the comma is moved one step forward. The inverse Bernoulli shift is simply moving the comma one step backward. Both the Bernoulli shift and its inverse are homeomorpisms. Thus, the Bernoulli shift is an automorphism.  The Bernoulli shift has sensitive dependence upon initial data - the signature of chaos. Take two close initial points $a$ and $b$ in $\Sg$, say $b$ is in the neighborhood $U_N(a)$ for a large $N$, but not in $U_{N+1}(a)$, that is
\[
a_n = b_n, \ \ |n|< N,
\]
but 
\[
\text{either } a_N \neq b_N \text{ or } a_{-N} \neq b_{-N} .
\]
After applying either the Bernoulli shift or its inverse $N$ times, the images are not inside $U_1$ neighborhood, that is, either $\chi^N(b)$ is not in $U_1(\chi^N(a))$ or $\chi^{-N}(b)$ is not in $U_1(\chi^{-N}(a))$. Thus no matter how close two different initial points are, after enough iterations of the Bernoulli shift, the images are far away. 

\section{van Veen - Kawahara transversal homoclinic orbit in plane Couette flow}

\begin{figure}[ht] 
\centering
\includegraphics[width=4.5in,height=3.0in]{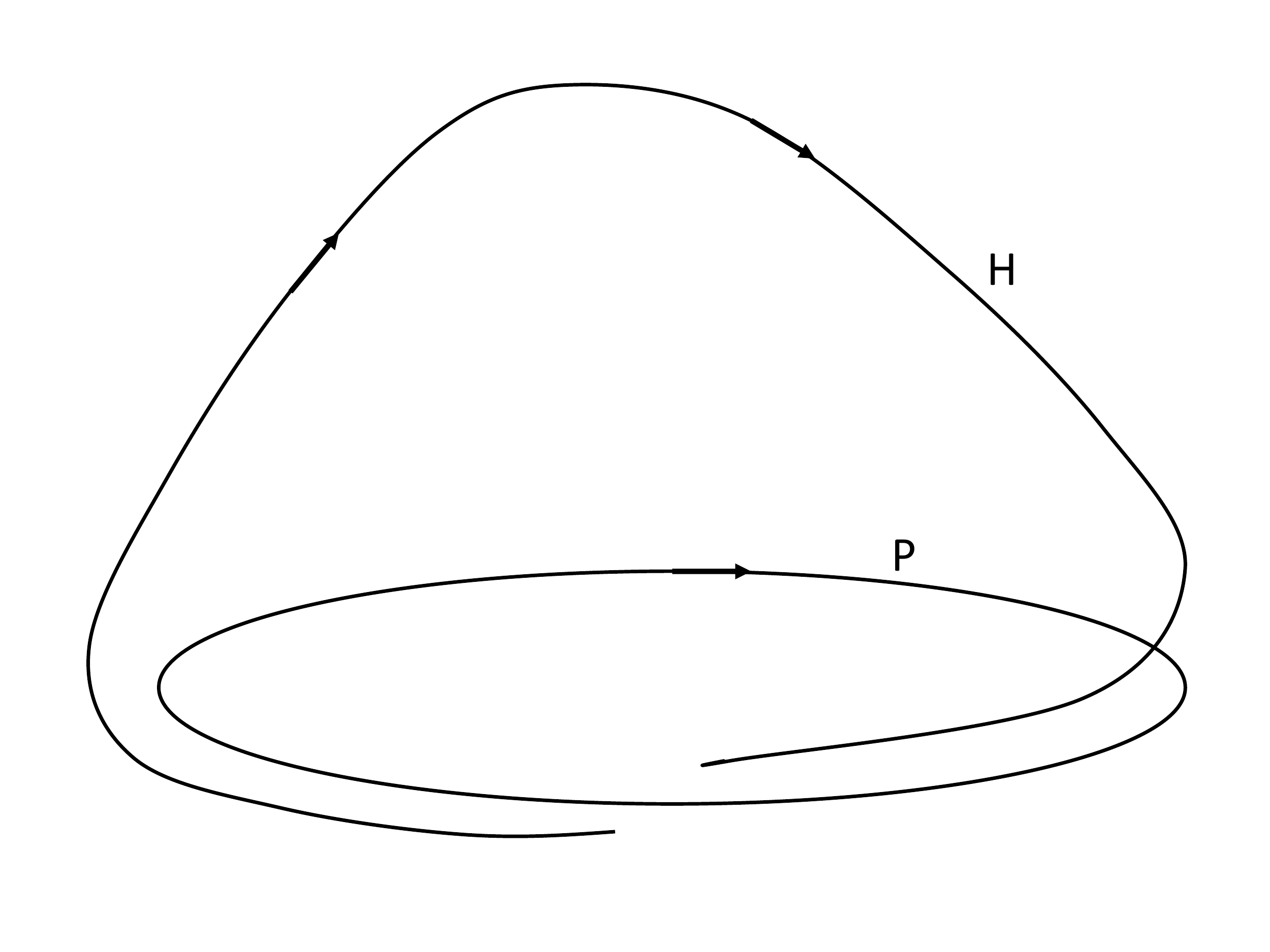}
\caption{Topological setup of the homoclinic orbit.}
\label{cou1}
\end{figure}

For the plane Couette flow at moderate Reynolds number ($\sim 400$), the topological setup of the homoclinic orbit numerically observed by L. van Veen and G. Kawahara \cite{VK11} can be depicted as in Figure \ref{cou1} where $H$ is the homoclinic orbit asymptotic, in both forward and backward time, to the limit cycle (isolated periodic orbit) $P$. The limit cycle $P$ has a two-dimensional unstable manifold and a one-codimensional stable manifold, which intersect transversally along the homoclinic orbit $H$. In \cite{VK11}, the limit cycle is very small, and the homoclinic orbit is very large. Nevertheless, the topology is the same as in Figure \ref{cou1}. In both forward and backward time, the homoclinic orbit just keeps wrapping around the limit cycle closer and closer. From the analysis of partial differential equations, the Poincar\'e return map is not well-defined on the Poincar\'e section transversal to the limit cycle. Thus one cannot reduce the problem of establishing chaos near the homoclinic orbit to that on the Poincar\'e section, instead we establish chaos directly in the original space \cite{Li03}.

\section{Pseudo-orbits}

\begin{figure}[ht] 
\centering
\includegraphics[width=4.5in,height=3.0in]{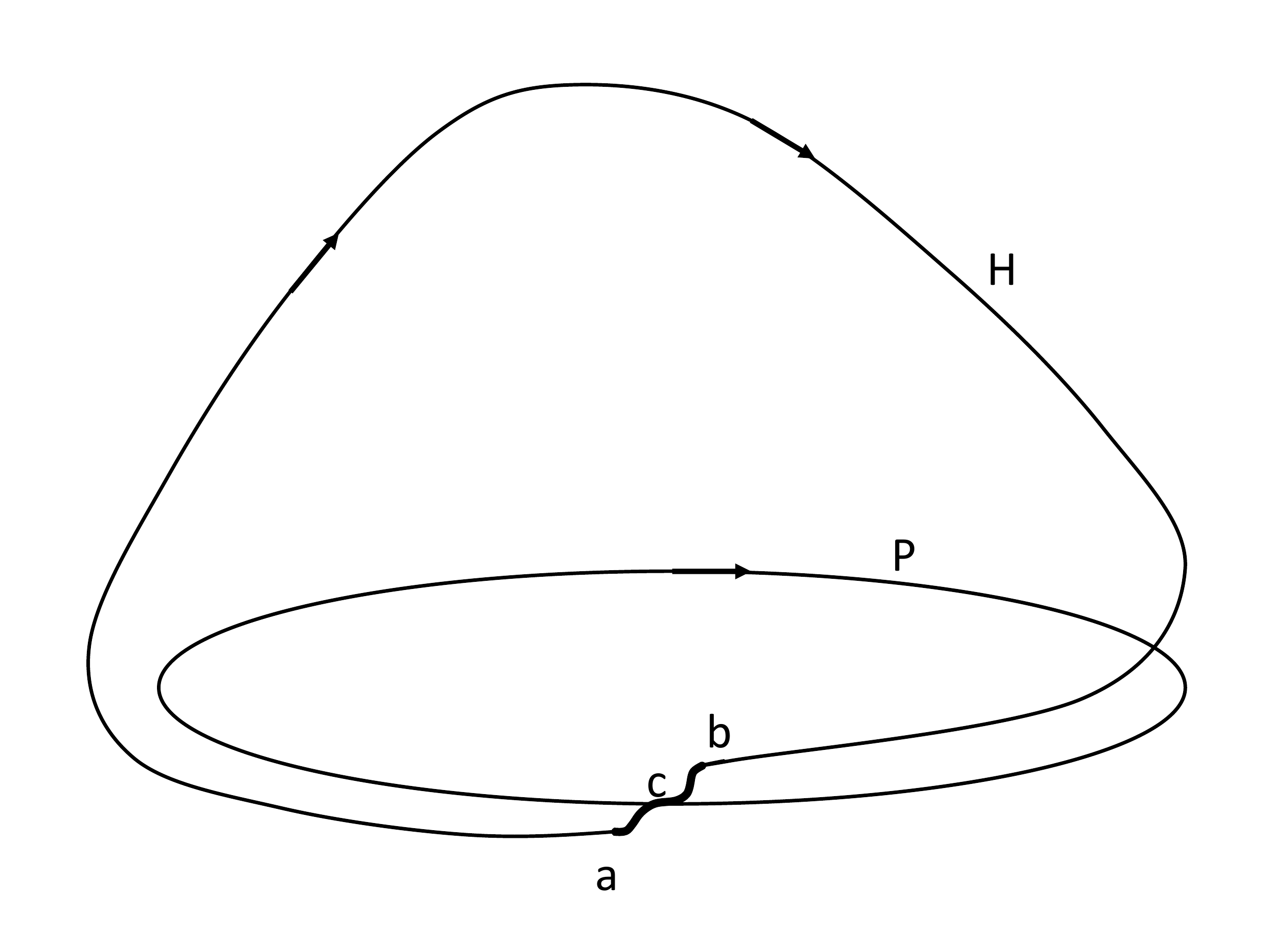}
\caption{Pseudo-orbit setup.}
\label{cou2}
\end{figure}

To establish the symbolic dynamics and chaos near the homoclinic orbit, we introduce pseudo-orbits \cite{Li03}.  Let $c$ be a point on the limit cycle $P$, and $\eta_0$ be the periodic orbit starting from $c$, that travels on $P$ many times (i.e. many copies of $P$).  We can follow the homoclinic orbit backward until it wraps around the limit cycle many times and reach a point $a$ that is close to the point $c$ on the limit cycle. Then the point $a$ is very close to the limit cycle. Also we can follow the homoclinic orbit forward until it wraps around the limit cycle many times and reach a point $b$ that is close to the point $c$ on the limit cycle. Again the point $b$ is very close to the limit cycle. See Figure \ref{cou2}. To make the illustration in Figure \ref{cou2} clearer, we did not draw the wrapping of the homoclinic orbit around the limit cycle. From $b$ to $a$, we put a mollifying connector that is tangent to the homoclinic orbit at $b$ and $a$, and tangent to the limit cycle at $c$. With this connector on, we get a closed loop denoted by $\eta_1$. While $\eta_1$ is not an orbit, it is a pseudo-orbit since the jump from $b$ to $a$ is small. The starting point and the end point of the pseudo-orbit $\eta_1$ is $c$.
\begin{figure}[ht] 
\centering
\includegraphics[width=4.5in,height=3.0in]{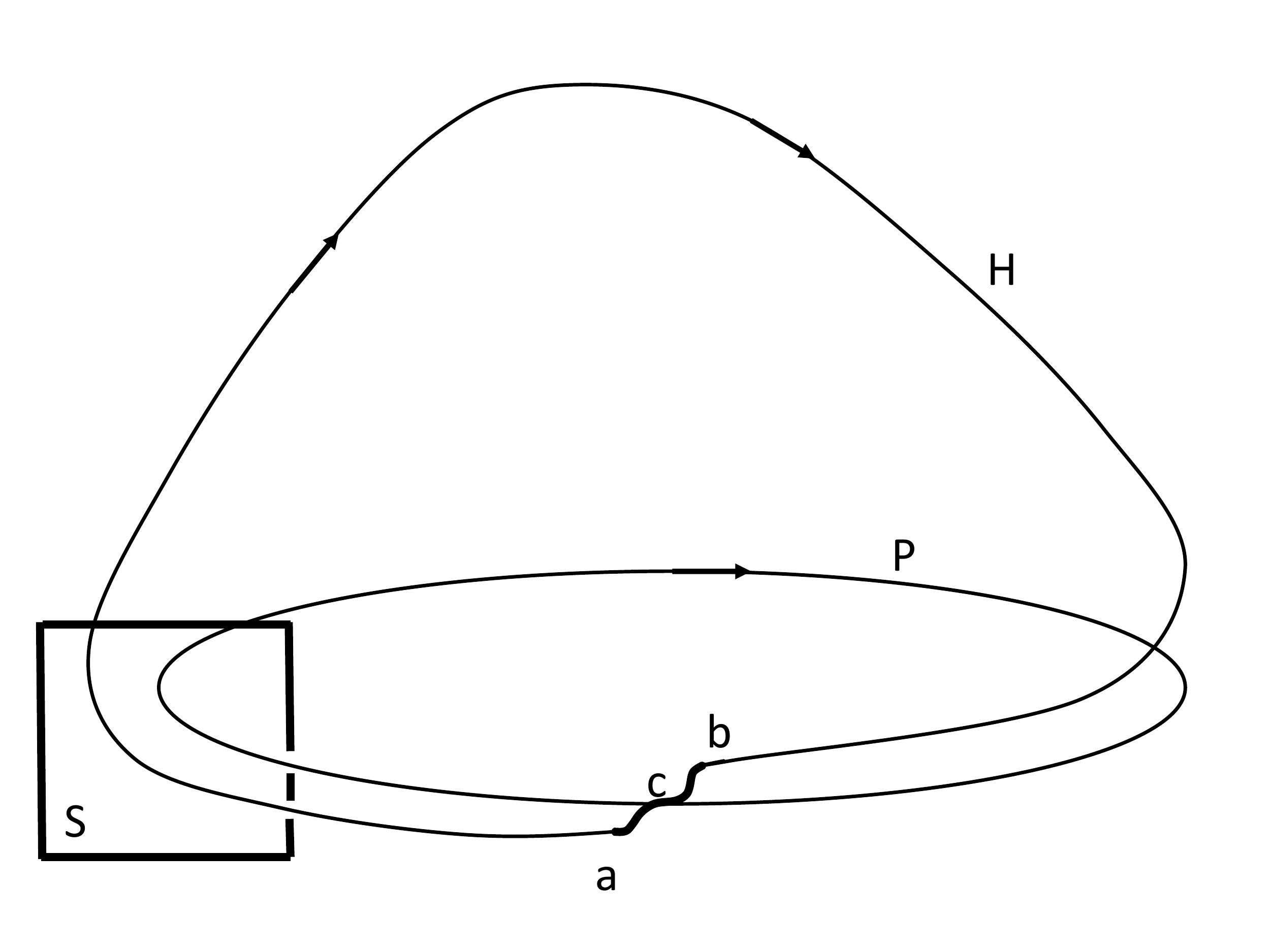}
\caption{Poincar\'e  section setup.}
\label{cou3}
\end{figure}
\begin{figure}[ht] 
\centering
\includegraphics[width=4.5in,height=3.0in]{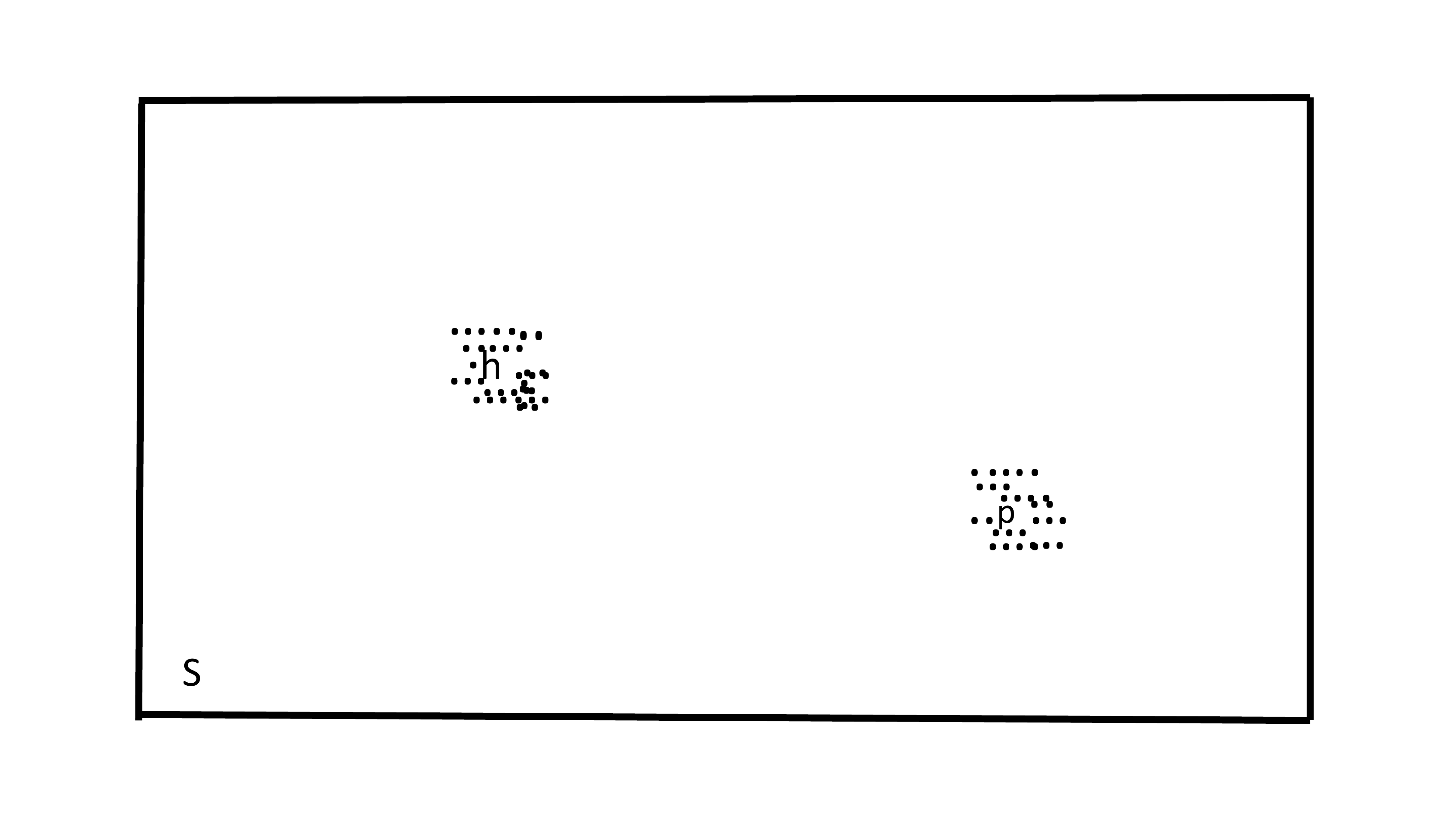}
\caption{Cantor subset of the Poincar\'e section.}
\label{cou4}
\end{figure}
To any doubly infinite binary sequence
\[
a= (\cdots a_{-2}a_{-1} a_0, a_1 a_2 \cdots ) ,
\]
where $a_n = 0$ or $1$ for all integers $n$, there corresponds a pseudo-orbit
\[
\eta_a = (\cdots \eta_{a_{-2}} \eta_{a_{-1}} \eta_{a_0}, \eta_{a_1} \eta_{a_2} \cdots  ).
\]
Note that the connecting point between $\eta_0$ and $\eta_1$ is $c$.

\section{Symbolic dynamics and chaos}

Through contraction map argument on the graphs of pseudo-orbits \cite{Li03}, we can show that to any doubly infinite binary sequence
\[
a= (\cdots a_{-2}a_{-1} a_0, a_1 a_2 \cdots ) ,
\]
where $a_n = 0$ or $1$ for all integers $n$, there corresponds an true orbit $O_a$ that is close to the pseudo-orbit $\eta_a$ \cite{Li03} (strictly speaking, the analysis in \cite{Li03} applies to all the Galerkin truncations of the Navier-Stokes equations, which are the systems that numerical simulations truly calculate. Due to analytical technicality, the analysis in \cite{Li03} does not directly apply to the full Navier-Stokes equations due to the singular perturbation nature of the Laplacian dissipative term). We call $O_a$ the shadowing orbit of the pseudo-orbit $\eta_a$. Thus the topological space $\Sg$ of all doubly infinite binary sequences is in one-to-one correspondence with the set $\Xi$ of these true orbits $O_a$'s. We put a Poincar\'e section $S$ in a location before the homoclinic orbit starts to wrap around the limit cycle in backward time (see Figure \ref{cou3}). Let $h$ be the intersection point of homoclinic orbit with the Poincar\'e section $S$ before the homoclinic orbit starts to wrap around the limit cycle in backward time. Let $p$ be the intersection point of limit cycle with the Poincar\'e section $S$. $h$ and $p$ are far away apart. See Figure \ref{cou4}. We designate $h$ the chosen intersection point of the loop $\eta_1$ with the Poincar\'e section $S$, and $p$ the chosen intersection point of the first copy of the limit cycle $P$ in the loop $\eta_0$. We designate the chosen intersection point of $\eta_a$ with the Poincar\'e section $S$ to be that of $\eta_{a_0}$. Finally, we designate the chosen intersection point of $\O_a$ with the Poincar\'e section $S$ to be the one that is close to that of $\eta_a$. See Figure \ref{cou4} for an illustration. Denote by $\La$ the set of intersection points of all $\O_a$'s with the Poincar\'e section $S$ for all doubly infinite binary sequences $a \in \Sg$. $\La$ has a Cantor type structure. The topology of $\La$ discussed above induces a topology on $\La$. For any point $o_* \in \La$ which is the intersection point of $\O_{a_*}$ with the Poincar\'e section $S$, the neighborhood $U_j(a_*)$ of $a_*$ induces a neighborhood $W_j(o_*)$ of $o_*$, which consists of the intersection points of all $\O_a$ with the Poincar\'e section $S$ for all $a \in U_j(a_*)$. The Poincar\'e return map $F$ is not defined on the Poincar\'e section $S$ as an analytical result of partial differential equations, but it is defined on $\La$. Defined on the topological space $\La$, the Poincar\'e return map $F$ is topologically conjugate (equivalent) to the Bernoulli shift $\chi$ defined on the topological space $\Sg$. That is, there is a homeomorphism $\phi : \Sg \ra \La$ such that the following diagram commutes:
\[
\begin{array}{ccc}
\Sg &\maprightu{\phi} & \Lambda\\
\mapdownl{\chi} & & \mapdownr{F}\\
\Sg & \maprightd{\phi} & \Lambda
\end{array} 
\]
Since the Bernoulli shift $\chi$ has sensitive dependence upon initial data in the topological space $\Sg$, the Poincar\'e return map $F$ also has sensitive dependence upon initial data in the topological space $\La$, that is, the dynamics of $F$ is chaotic. If we choose two different nearby points in $\La$, then after enough iterations or inverse iterations of $F$, the image of one point is close to $p$ and that of the other is close to $h$ (note that $p$ and $h$ are far away apart).

\end{document}